\documentclass[a4paper,11pt]{article}
\usepackage{jinstpub} % for details on the use of the package, please see the JINST-author-manual
\usepackage{lineno}
\usepackage{etoolbox}
\usepackage{hyperref}
\usepackage{url}
\usepackage{booktabs}
\usepackage{microtype}
\usepackage{natbib}
\usepackage{float}
\usepackage{subfigure}

\title{Refine Neutrino Events Reconstruction with BEiT-3}

\author{Chen Li,}
\author{Hao Cai,}
\author[1]{and Xianyang Jiang\note{Corresponding author. E-mail: \url{jiang@whu.edu.cn}}}
\affiliation{School of Physics and Technology, Wuhan University,\\
Wuhan, China}

\abstract{Neutrino Events Reconstruction has always been crucial for IceCube Neutrino Observatory. In the Kaggle competition ``IceCube -- Neutrinos in Deep Ice'', many solutions use Transformer. We present ISeeCube, a pure Transformer model based on $\texttt{TorchScale}$ (the backbone of BEiT-3). When having relatively same amount of total trainable parameters, our model outperforms the  $2^{\textrm{nd}}$ place solution. By using $\texttt{TorchScale}$, the lines of code drop sharply by about 80\% and a lot of new methods can be tested by simply adjusting configs. We compared two fundamental models for predictions on a continuous space, regression and classification, trained with MSE Loss and CE Loss respectively. We also propose a new metric, overlap ratio, to evaluate the performance of the model. Since the model is simple enough, it has the potential to be used for more purposes such as energy reconstruction, and many new methods such as combining it with $\texttt{GraphNeT}$ can be tested more easily. The code and pretrained models are available at \url{https://github.com/chenlinear/ISeeCube}.}


\arxivnumber{2308.13285} % Only if you have one

\begin{document}
\maketitle{}
\flushbottom

\section{Introduction}
\label{sec:headings1}

IceCube is a neutrino observatory that studies the most extreme phenomena in the universe by observing neutrinos \cite{aartsen1612jinst}. Examples include Gamma-ray Bursts, Active Galactic Nuclei, and recently discovered neutrinos emitted by the Milky Way \cite{icecube2023observation}. Neutrino Events Reconstruction such as direction reconstruction and energy reconstruction are crucial in the observation of neutrinos. Traditionally, to perform the reconstruction, we would use maximum likelihood methods or Machine Learning models that mostly based on Convolutional Neural Networks (CNNs) \cite{abbasi2021convolutional, eller2023flexible, aiello2020event, hunnefeld2021combining, peterson20232d}, or Graph Neural Networks (GNNs), \cite{abbasi2022graph, sogaard2022graphnet}.

In order to encourage the development of innovative solutions to improve the accuracy and efficiency of Neutrino Events Reconstruction, the Kaggle competition ``IceCube -- Neutrinos in Deep Ice'' \cite{eller2023public} was held between January 19, 2023 and April 19, 2023. Most top solutions use Transformer or $\texttt{GraphNeT}$ integrated with Transformer. Transformer is traditionally used for Natural Language Processing \cite{vaswani2017attention}, and then Computer Vision \cite{dosovitskiy2020image} by introducing the idea of class token (cls token) and only using the encoder part of the full Transformer, which is called Vision Transformer (ViT). Transformer is powerful that it can learn the position information of a sentence or a picture.

In particular, the $2^{\textrm{nd}}$ place solution IceMix uses BEiT v2 \cite{peng2022beit} block as encoder. BEiT v2 uses semantic-rich visual tokenizer as the reconstruction target for masked prediction in Computer Vision. However, current models are heavy and the benefits of each other is hard to be gathered.

BEiT-3 \cite{wang2022image} is a general-purpose multimodal foundation model based on Magneto \cite{wang2022foundation}. And it can be used by importing $\texttt{TorchScale}$ \cite{ma2022torchscale}, which is an open-source toolkit that enables scaling Transformers both efficiently and effectively. ISeeCube is a pure Transformer model based on $\texttt{TorchScale}$. A pure Transformer structure has been proven powerful in many Machine Learning missions, and in Neutrino Events Reconstruction through this competition. And by utilizing the highly integrated $\texttt{TorchScale}$, it is easier for refining, maintaining and testing.

For upcoming neutrino observatories such as TRIDENT \cite{ye2023multi} and HUNT \cite{huang2023proposal}, considering that (a) they also observe astrophysical neutrinos and (b) their structure is close to IceCube that they are also composed of several strings of Digital Optical Modules (DOMs), there's a great chance that ISeeCube can also be used for these observatories.

The paper is structured as follows: Section \ref{sec:headings2} summarize Neutrino Events Reconstruction methods and some of the top solutions in this Kaggle competition. We present ISeeCube as a pure Transformer structure for Neutrino Events Reconstruction in section~\ref{sec:headings3}, and compare regression models and classification models. Followed by section~\ref{sec:headings4} we describe the dataset and the training process, and we evaluate the azimuthal and zenithal distribution of the prediction made by regression and classification models. In section~\ref{sec:headings5} we discuss the potential of ISeeCube and challenges in other missions such as energy reconstruction and cascade/track event classification.

\section{Related Work}
\label{sec:headings2}

Previously, likelihood method \cite{bellenghi2023extending, glusenkamp2023conditional} is widely used for Neutrino Events Reconstruction. Also, there are Machine Learning models based on CNNs \cite{abbasi2021convolutional, eller2023flexible, aiello2020event, hunnefeld2021combining, peterson20232d} or GNNs \cite{abbasi2022graph, sogaard2022graphnet}. These two Neural Networks (NNs) represent two different approaches of data representation, see section~\ref{sec:headings31}. Notably, neutrinos from the Milky Way are detected \cite{icecube2023observation} by first using CNN \cite{abbasi2021convolutional} to classify cascade events from all the events, and then using hybrid reconstruction method combined with a maximum likelihood estimation \cite{hunnefeld2021combining}.

In private leaderboard of the Kaggle competition \footnote{\url{https://www.kaggle.com/competitions/icecube-neutrinos-in-deep-ice/leaderboard}}, these top solutions are representative:

\begin{itemize}
    \item $1^{\textrm{st}}$: It contains 3-4 layers and the number of total parameters is only 6M (one of the smallest top solutions). Each of these layer is composed of an \texttt{EdgeConv} layer and a Transformer Encoder layer sequentially.
    \item $2^{\textrm{nd}}$: The Transformer part is based on BEiT v2 \cite{peng2022beit} block. Also, it uses class token and introduces relative spacetime interval bias as Relative Positional Embedding (RPE) \cite{shaw2018self}. It has two Encoders, the first Encoder uses RPE and the second does not. This solution shows that, with right set of parameters for RPE, embedding dimensions and number of layers, a pure Transformer structure works as well. RPE is close to the mechanism that is used in GNNs to capture the relationship of nodes in a sequence. This solution also uses GNNs as the first layer of the model to further improve the score.
    \item $3^{\textrm{rd}}$: It's a pure Transformer structure, which has two classification heads (for azimuth and zenith) with 128 bins each. Since it's a classification mission, a Cross Entropy Loss is better than Von Mises-Fisher Loss \cite{kumar2018mises}.
    \item $5^{\textrm{th}}$: The $\texttt{GraphNeT}$ part uses LayerNorm as the activation function instead of BatchNorm1d. The Transformer part has two classification heads, each containing $256$ bins for azimuth and zenith. The Transformer consists of $8$ layers, with a model size of $512$ and $8$ attention heads. Additionally, a $3$-layer Bidirectional LSTM model selector is employed to choose between $\texttt{GraphNeT}$ and Transformer. Combining multiple individual models to solve a given problem is called ``ensemble'' method.
\end{itemize}

Besides GNNs, which wins the Early Sharing Prize and is provided as an example in this competition, some solutions also use Recurrent Neural Networks (RNNs), which take into account of time information and treat neutrino events as time sequences. Long Short-Term Memory (LSTM) and Gate Recurrent Unit (GRU) are designed to deal with long-range dependencies. But as the original Transformer paper \cite{vaswani2017attention} stated, a pure Transformer model instead of RNNs can be used to solve the same problem and reach better score.

\begin{table}[H]
\centering
\footnotesize
\caption{Machine Learning methods in some of the Kaggle solutions.}
\begin{tabular}{lccccc}
\hline
 & Transformer & GNN & RNN & Loss & Ensemble \\
\hline
$1^{\textrm{st}}$ & $\checkmark$ & EdgeConv &  & VMF (custom) & $\checkmark$ \\
$2^{\textrm{nd}}$ & BEiT v2 & EdgeConv (optional) &  & VMF & $\checkmark$ \\
$3^{\textrm{rd}}$ & NanoGPT &  &  & CE & $\checkmark$ \\
$5^{\textrm{th}}$ & $\checkmark$ & GraphNeT & LSTM & CE & $\checkmark$ \\
$6^{\textrm{th}}$ & $\checkmark$ & $\checkmark$ & LSTM & CE (custom) & $\checkmark$ \\
$8^{\textrm{th}}$ &  & GraphNeT, GPS, GravNet &  & VMF, CosineSimilarity &  \\
$10^{\textrm{th}}$ &  &  & LSTM & CE, VMF &  \\
$11^{\textrm{th}}$ & $\checkmark$ & $\checkmark$ & GRU &  & $\checkmark$ \\
$12^{\textrm{th}}$ &  & DynEdge & LSTM &  & $\checkmark$ \\
$13^{\textrm{th}}$ &  & GraphNeT &  &  & $\checkmark$ \\
$14^{\textrm{th}}$ & $\checkmark$ &  & BGRU &  &  \\
$15^{\textrm{th}}$ &  & DynEdge &  & VMF &  \\
$17^{\textrm{th}}$ &  & DynEdge & LSTM &  & $\checkmark$ \\
$20^{\textrm{th}}$ &  &  & LSTM &  & $\checkmark$ \\
\hline
\end{tabular}
\label{tab:table1}
\end{table}

Table~\ref{tab:table1} summarizes some of the top solutions in the Kaggle competition \cite{bukhari2023icecube}. ``$\checkmark$'' means the architecture is used in the solution, and a specific name represents the specific model. ``Ensemble'' means whether the final result is a ensemble of several different models by some weighed mechanism.

In summary, Transformer architecture can work really well for having the ability to capture the information of both graphs \cite{joshi2020transformers} and time sequences. What's more, since the total number of nodes in a graph is smaller than vocabulary of a language, ViT is better than Transformer that an event is closer to a picture than sentence.

As for loss function, Von Mises-Fisher Loss \cite{kumar2018mises} has the advantage that the first Bessel function is the solution for partial differential equations in a cylinder. And IceCube is constructed by several strings, meaning that all the detectors on the same string has the same $x$ and $y$ coordinate, which makes its shape closer to a cylinder. This loss function finds the internal properties of IceCube. However, some solutions indicate that Cross Entropy Loss is better than Von Mises-Fisher Loss. A possible explanation is that these solutions divide the azimuth and zenith into several bins and then predict which bin is the most likely, thus it can work really well with Cross Entropy Loss, which is normally used for classification missions.

\section{ISeeCube}
\label{sec:headings3}

We noticed that in the $2^{\textrm{nd}}$ place solution, new layers and blocks have to be re-written to apply Special Relativity bias and the structure of the model is not clear (see section~\ref{sec:headings32}). Also, MSE Loss and CE Loss are simpler than VMF Loss (see section~\ref{sec:headings34}). In order to enhance structural clarity and simplicity, reduce lines of code, and investigate the feasibility of implementing Transformers and their upgrades for Neutrino Events Reconstruction, we made several enhancements to the $2^{\textrm{nd}}$ place solution, see table~\ref{tab:table2}. The code for embedding (see section~\ref{sec:headings31}) and data-loading (see section~\ref{sec:headings41}) are inherited from the $2^{\textrm{nd}}$ place solution. The rest of the code are re-written.

\begin{figure}[t]
	\centering
	%\subfigbottomskip=2pt
	\subfigcapskip=-5pt
	\subfigure[ISeeCube]{
		  \includegraphics[width=0.6\linewidth]{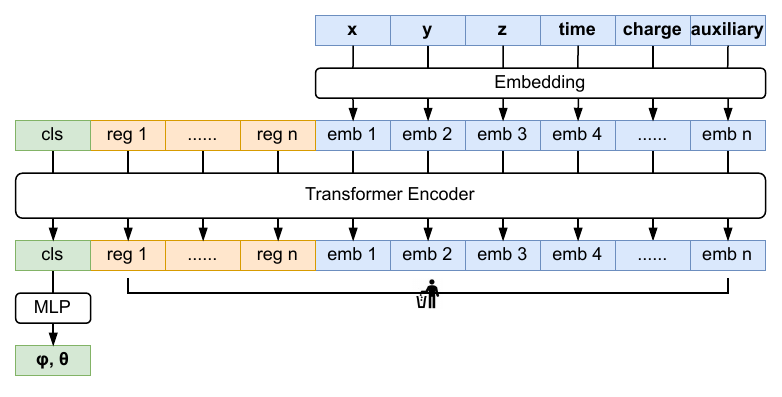}}
        \hspace{0.1em}
	\subfigure[Transformer Encoder]{
            \includegraphics[width=0.3\linewidth]{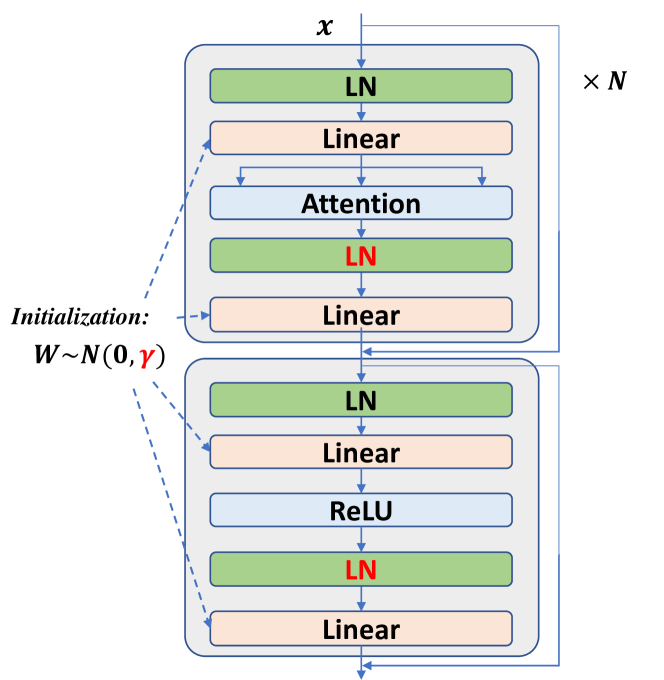}}
        %\captionsetup{width=0.85\textwidth}
	\caption{The structure of ISeeCube. (a) shows how raw data is embedded then put into the encoder, the class token is taken out and transformed into azimuth and zenith as the result, while the register tokens and embedded data are removed. ``n'' in ``emb n'' corresponds to the number of pulses in an event, while ``n'' in ``reg n'' is a rather arbitrary number. (b) shows the detail of each layer in the Transformer Encoder (originally figure 2 (a) in Magneto \cite{wang2022foundation}).}
        \label{fig:fig1}
\end{figure}

\begin{table}[H]
\centering
\footnotesize
\caption{Differences and commonalities between the $2^{\textrm{nd}}$ place solution and ISeeCube.}
\begin{tabular}{l|c|c}
\hline
 & $2^{\textrm{nd}}$ place solution & ISeeCube \\
\hline
Embedding & \multicolumn{2}{c}{Sinusoidal Positional Embedding} \\
\hline
Transformer Encoder & BEiT v2 & BEiT-3 ($\texttt{TorchScale}$) \\
\hline
Number of Encoders & 2 & 1 or 2 \\
\hline
RPE & $\checkmark$, Special Relativity Bias & $\checkmark$ \\
\hline
Token & Class Token & Class Token, Register Tokens \\
\hline
Loss & VMF & MSE or CE \\
\hline
Data-Loading & \multicolumn{2}{c}{Chunk-Based Loading, Batch-Based Loading} \\
\hline
\end{tabular}
\label{tab:table2}
\end{table}

We present a pure Transformer model for direction reconstruction: ISeeCube. ISeeCube has the potential to be extended to other tasks, such as energy reconstruction, cascade/track event classification, etc. In this section, we will elaborate the structure of ISeeCube:

\begin{itemize}
    \item In section~\ref{sec:headings31}, we summarize previous data representation of neutrino events for CNNs and GNNs, and conclude that \textbf{Embedding} in Transformer is a intuitive derivation of the previous methods.
    \item In section~\ref{sec:headings32} we show that by using $\texttt{TorchScale}$, the lines of code drop sharply. We show that ViT-like model is capable of learning the position and time information of neutrino events, and future new models or upgrades on \textbf{Transformer Encoder} and ViT has the potential to be used on Neutrino Events Reconstruction. Simple methods work really well, such as \textbf{Sinusoidal Positional Embedding}, \textbf{Relative Position Embedding} (RPE) \cite{raffel2020exploring}, \textbf{Class Token} and \textbf{Register Tokens} \cite{darcet2023vision}.
    \item In section~\ref{sec:headings33} we compare regression and classification models by using different \textbf{Multilayer Perceptrons} (MLPs) on the class token.
    \item In section~\ref{sec:headings34} we compare \textbf{Mean Squared Error} (MSE) Loss and \textbf{Cross Entropy} (CE) Loss used in regression and classification models, respectively. We also discuss the usage of the Kaggle competition metric.
\end{itemize}

\subsection{Embedding: Data Representation}
\label{sec:headings31}

Machine Learning models in Computer Vision mostly focus on images that satisfy the following properties:

\begin{itemize}
    \item Image $\mathbf{x}$ has the shape $\mathbb{R}^{H \times W \times C}$. This means that (a) the image is 2-dimensional, thus $H$ (Height) and $W$ (Width), and (b) the image has a grey channel ($C=1$), or RGB channels ($C=3$).
    \item Image is compactly arranged, meaning that the distance between each pixels is the same. And in most cases, zero.
\end{itemize}

\begin{figure}[H]
	\centering
        \includegraphics[width=0.45\linewidth]{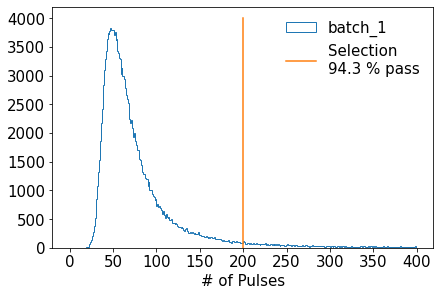}
        %\captionsetup{width=0.85\textwidth}
	\caption{In batch one, 94.3\% of events have less or equal to 200 pulses. Most events have about $50$ pulses. This figure is originally used in \url{https://www.kaggle.com/code/rasmusrse/graphnet-example}.}
	\label{fig:fig2}
\end{figure}

On the other hand, in IceCube, the neutrino events satisfy the following properties:

\begin{itemize}
    \item Intuitively, considering all the DOMs (no matter detect photons in this particular event or not), neutrino event can be represented as a 3-dimensional ``image'' $\mathbb{R}^{H \times W \times L \times P}$. This means that (a) neutrino event is 3-dimensional, thus $H$ (Height), $W$ (Width) and $L$ (Length), and (b) each pulse (detection made by DOM) has several properties, in our case time, charge and auxiliary ($P=3$).
    \item Neutrino event is loosely arranged, meaning that (a) the distance between each DOM is not the same, and (b) not all the DOMs detect photons in an event, see figure~\ref{fig:fig2}.
\end{itemize}

One of the problems of this intuitive representation might be computation because (a) this matrix is much larger than images in traditional Machine Learning methods and (b) large part of the matrix is empty (sparse matrix). Thus in many previous studies the intuitive representation is not used. In the rest of this section we will review some of the methods used in CNNs, GNNs and Transformers.

\textbf{CNNs}: In some CNNs models \cite{peterson20232d, abratenko2021semantic}, neutrino event is represented as $\mathbb{R}^{S \times N \times T}$, where $S$ is the number of strings, $N$ is the number of DOMs on each string and $T$ is time segments. Time segments is the total time divided by time periods. And the value in this tensor represents charge. This tensor is then put into a CNN. For example, in \cite{abratenko2021semantic} the CNN is based on ResNet \cite{he2016deep}.

\textbf{GNNs}: In $\texttt{GraphNeT}$ and many other GNN-based methods, the neutrino event is represented as $\mathbb{R}^{N \times P}$, where $N$ is the number of DOMs that detect photons in a particular event, $P$ is the number of properties a pulse has, in our cases position ($x$, $y$, $z$), time, charge and auxiliary ($P=6$). This matrix is then put into a GNN. For example, in $\texttt{GraphNeT}$, $\texttt{TAGConv}$ and $\texttt{EdgeConv}$ from the package $\texttt{PyTorch Geometric}$ are used.

\textbf{Transformers}:  Same as the GNN method, the neutrino event is also represented as $\mathbb{R}^{N \times P}$. Firstly, when the total number of pulses is bigger than the given $N$, we would choose auxiliary False first, then True; when the total number of pulses is smaller than $N$, the empty space will be filled with the position of the first DOM (the choice of the first DOM is rather arbitrary) and time, charge, auxiliary would be zero. For more information on the dataset and how to normalize the event, see section~\ref{sec:headings41}. Then \textbf{Sinusoidal Positional Embedding} is used. In the original Transformer \cite{vaswani2017attention}, Sinusoidal Positional Embedding assigns a unique fixed vector to each position in the input sequence. These vectors are computed using sine and cosine functions with different frequencies. In neutrino events it is introduced for the same purpose, projecting $\mathbb{R}^{N \times P}$ to a higher dimension. In our case, for example, one of the properties $x$ (shape $\mathbb{R}^{N \times 1}$), is projected to $\mathbb{R}^{N \times N}$ by:
\begin{equation}
\begin{aligned}
PE_{(i, j)} &= \sin \left( \frac{pos(i)}{10000^{j/N}} \right), j = 0, 1, \cdots , \frac{N}{2} - 1 \\ PE_{(i, j)} &= \cos \left( \frac{pos(i)}{10000^{j/N}} \right) , j = \frac{N}{2}, \frac{N}{2}+1, \cdots , N-1
\end{aligned}
\end{equation}
where $pos(i)$ is each element in $x$. Then the rest of the properties will go through the same process and all the embedded elements will be concatenated together ($x$, $y$, $z$ will be concatenated first, then the rest of the properties). Then we use a MLP to produce the final embedded event as $\mathbb{R}^{N \times d_{model}}$. This is the input of the Transformer Encoder. 

As we can see, Embedding used in Transformers is an intuitive upgrade of GNNs. Also, normally, when using higher embedding dimension, we will get better performance, in which way we can weigh the balance between computation and performance more easily. While in GNNs, different structures have to be tested multiple times before we can find the most suitable structure for specific model and goal.

\subsection{Transformer Encoder}
\label{sec:headings32}

We present an easy-to-use and simple backbone structure with lines of code dropping sharply by about 80\%. $\texttt{TorchScale}$ is portable and easy to maintain:

\begin{itemize}
    \item Out-of-the-box: An encoder structure can be imported by few lines of code. A lot of methods can be tested by adjusting the config. For example, Relative Position Embedding (RPE). While in the $2^{\textrm{nd}}$ place solution, the layers for encoder have be rewritten. Specifically, besides \texttt{class Block} from BEiT v2, several new classes \texttt{Attention\_rel}, \texttt{Block\_rel}, \texttt{Rel\_ds} and \texttt{LocalBlock} have to be modified or introduced to apply Special Relativity bias. \footnote{Please refer to \url{https://github.com/DrHB/icecube-2nd-place/blob/main/src/models.py} and \url{https://github.com/microsoft/unilm/blob/master/beit2/modeling_finetune.py}.}
    \item Only rely on few dependencies including $\texttt{PyTorch}$, $\texttt{timm}$ and $\texttt{fairscale}$, while GNNs rely on $\texttt{PyTorch}$, $\texttt{PyTorch Cluster}$, $\texttt{PyTorch Scatter}$, $\texttt{PyTorch Sparse}$ and $\texttt{PyTorch Geometric}$.
    \item In Feed Forward Network (FFN), A sub\_LayerNorm (sub\_LN) is added to stabilize models with deeper layers. The default activation function in FFN is now GELUs \cite{hendrycks2016gaussian} rather than ReLUs, which combines dropout zoneout and ReLUs to alleviate the vanishing gradient problem. And a dropout layer is added to alleviate the same problem.
\end{itemize}

\textbf{Relative Position Embedding} (RPE) \cite{raffel2020exploring} is presented as a new method to learn the relative position information. In the $2^{\textrm{nd}}$ place solution, RPE is added by introducing Special Relativity bias and writing new layers and blocks with this bias. In $\texttt{TorchScale}$, RPE can be added easily by adjusting few parameters, though without Special Relativity bias.

In ViT \cite{dosovitskiy2020image}, \textbf{Class Token} is concatenated to the embedding of the picture to learn the position information. In ISeeCube, Class Token is concatenated to the embedding of a neutrino event for the same purpose, to learn the position information of neutrino events.

\textbf{Register Tokens} \cite{darcet2023vision} are used for smoother feature maps and attention maps for downstream visual processing when finetuning ViT models. In ISeeCube, Register Tokens are concatenated to the embedding of a neutrino event for the same purpose. For image classification, the predicted class corresponds to the index with the largest probability; for Neutrino Events Reconstruction, the model directly predict the value itself. Thus smoother feature maps and attention maps are really helpful. Also, in terms of memory usage, we notice that when the number of class token is $1$, the number of register tokens is $3$, and we pick $196$ pulses in an event, the total sequence length been put into the encoder is $200$. The ``warp size'' of a CUDA core is the number of threads in a warp executed simultaneously on a CUDA core, which is normally $32$. When multiple threads within a warp access adjacent memory locations, the GPU can coalesce these memory transactions into a single memory transaction, resulting in improved memory throughput. $200$ is a multiple of $8$, which is more CUDA-friendly than the case with only class token, $197$, which is a prime number. 

\subsection{Multilayer Perceptron: Regression and Classification}
\label{sec:headings33}

For a continuous space, regression and classification are commonly used methods to predict the output of a neural network. We use different Multilayer Perceptrons (MLPs) as the final layer to build these two models:

\begin{itemize}
    \item Regression: The model predicts the output itself. The output is produced by a MLP that results with a tensor with shape $[B, 3]$, where $B$ is mini-batch size and 3 represents $x_{pred}$, $y_{pred}$, $z_{pred}$.
    \item Classification: We classify the continuous space into several bins then the model predicts which bins does the prediction belongs to. The predicted azimuth is produced by a MLP that results with a tensor with shape $[B, N_{bins}]$, where for azimuth, we use a MLP head with $N_{bins} = 128$; for zenith, we use another MLP head with $N_{bins} = 64$, so that the width for both bins is $\frac{\pi}{64}$. Then we use softmax to choose the largest index, which corresponds to an angle.
\end{itemize}

We trained three models (The hyper-parameter is summarized in table~\ref{tab:table4}. For detailed training configuration and loss curve, please see section~\ref{sec:headings42}.):

\begin{itemize}
    \item Model S-RegA is a regression model using the structure of figure~\ref{fig:fig1}.
    \item Model S-RegB is a regression model using similar structure, replacing one encoder with two encoders with and without Relative Position Bias, and residual connections are used between these two encoders.
    \item Model S-ClsB is a classification model using the same structure as model S-RegB.
\end{itemize}

We tried to train a classification model using the same structure as model S-RegA, but the loss does not go down. This is one of the disadvantages of classification model that they tend to be more unstable.

\subsection{Loss}
\label{sec:headings34}

We trained the regression model using Mean Squared Error (MSE) Loss, and the classification
model using Cross Entropy (CE) Loss. The Competition Metric is used as metric in the validation
process and as loss function in the final training epochs.

For regression models, \textbf{Mean Squared Error} (MSE) Loss calculates mean squared error between each element in the prediction $x$ and true value $y$: 
\begin{equation}
\textrm{MSE} \left( x, y \right) =\frac{1}{n} \sum_{i} \left( x_i-y_i \right)^2
\end{equation}
where $x_i$ and $y_i$ can be in any shape. In our case, $x_i$ and $y_i$ are both one-dimensional array with length $3$ (representing $x_{\text{reco}}$, $y_{\text{reco}}$, $z_{\text{reco}}$ and $x_{\text{true}}$, $y_{\text{true}}$, $z_{\text{true}}$ respectively). Also, to prevent exploding gradients, it could be replaced with Smooth L1 Loss \cite{girshick2015fast}.

As for classification models, following the Kaggle solutions stated in section~\ref{sec:headings2}, we use another commonly used loss function \textbf{Cross Entropy} (CE) Loss:
\begin{equation}
\textrm{CE} \left( x, y \right) =\frac{1}{n} \sum_i x_i \log{y_i}
\end{equation}
where $x_i$ is every item in the predicted probability and $y_i$ is the true probability index. We perform CE Loss on two directions, azimuth and zenith, respectively. And for each angle, $x$ and $y$ are one-dimensional array with length $N_{bins}$.

\textbf{Mean Angular Error}, used as the Kaggle competition metric \cite{eller2023public}, is the mean arccosine of the inner product of the true angle and the predicted angle: 
\begin{equation}
\Psi = \arccos (\sin{\vartheta_\text{true}}\sin{\vartheta_\text{reco}}(\cos{{\varphi_\text{ture}}}\cos{{\varphi_\text{reco}}}+\sin{{\varphi_\text{ture}}}\sin{{\varphi_\text{reco}}})+\cos{\vartheta_\text{true}}\cos{\vartheta_\text{reco}})
\end{equation}
Random guessing will yield a result of $\frac{\pi}{2}$, when the two vectors are orthogonal.

\section{Experiments}
\label{sec:headings4}

In section~\ref{sec:headings41} we summarize some of the most important aspects of the dataset and the normalization of neutrino events. In section~\ref{sec:headings42} we show the learning rate schedule, loss curve and performance of our models, followed by section~\ref{sec:headings43} we further explore the performance of our models by comparing azimuthal and zenithal distribution, and propose a new metric: Overlap Ratio.

\subsection{Dataset}
\label{sec:headings41}

Each event in the dataset is composed of a series of pulses. Each pulse has $6$ properties: DOM ID (which corresponds to position $x$, $y$, $z$), time, charge and auxiliary.

Most neutrino events have less or equal to $200$ pulses, and in most Kaggle solutions the dimension of the model is less than $200$, meaning that if the number of pulses is bigger than the dimension, some of the pulses will be discarded. A common used method is choosing auxiliary. Auxiliary determines the readout mode (LED current waveform) \cite{aartsen1612jinst}. When $\textrm{auxiliary} = \textrm{False}$, it means that at least one neighboring DOM on the same string also records a signal within $1 \mu s$. Thus normally False would give us a better result in the shape of the event, thus better direction reconstruction performance. This is why during the embedding process, when the number of pulses is bigger than the dimension of the model, pulses with auxiliary $\textrm{False}$ are chosen first, then $\textrm{True}$.

The biggest challenge of the dataset is its bigness (more than $138$ million events distributed in $660$ batches with the size of over $100$ GBs), so we use the same chunk-based loading method used in the $2^{\textrm{nd}}$ place solution, meaning that every epoch is trained on $654$ batches divided into $8$ sub-epochs, see section~\ref{sec:headings42}. The advantage of this chunk-based method is that now we can train the model on the entire dataset with relatively low memory cost. However it may have a disadvantage that a random sampling can only sampling across the sub-set of the entire dataset, which may leads to bias. For inference, we used the same batch-based loading method, meaning that we only load the desired batch. This is more flexible.

\begin{figure}[H]
	\centering
        \includegraphics[width=0.85\linewidth]{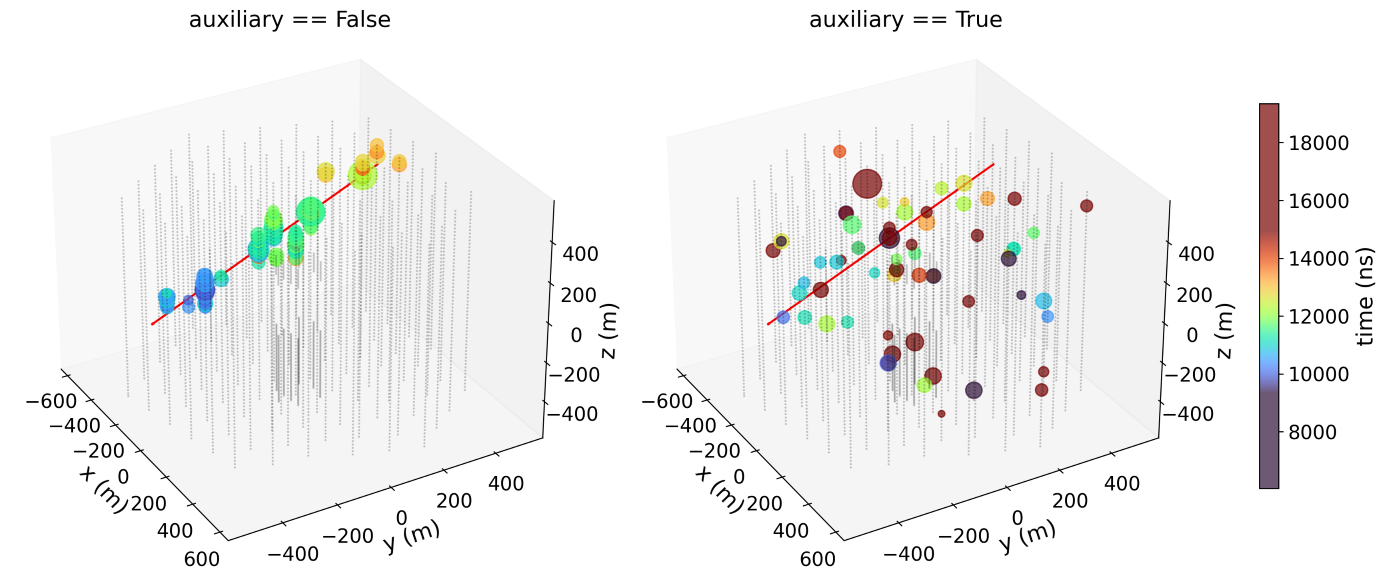}
        %\captionsetup{width=0.85\textwidth}
	\caption{This is an example event where auxiliary $\textrm{False}$ and $\textrm{True}$ are shown separately. It is intuitive that False will get better result in terms of shape thus direction reconstruction. This figure is originally used in \cite{eller2023public}.}
	\label{fig:fig3}
\end{figure}

\subsection{Training and Performance}
\label{sec:headings42}

For comparison, the hyper-parameters for $2^{\textrm{nd}}$ place solution and ISeeCube are shown in table~\ref{tab:table3} and table~\ref{tab:table4}. In order to better compare the performance of different Embedding methods and Transformer Encoders, the models that combines Transformers and GNNs in the $2^{\textrm{nd}}$ place solution is not included in table~\ref{tab:table3}.

\begin{table}[H]
\centering
\footnotesize
\caption{Hyper-parameters and score of $2^{\textrm{nd}}$ place solution without the models that are a combination of Transformer and GNN. ``T'', ``S'' and ``B'' refers to ``Tiny'', ``Small'' and ``Base''. ``d32'' or ``d64'' means the dimension of the head is 32 or 64. CV512 means the number of pulses in an event is 512 when inferencing on the validaiton dataset (batch 655 $\sim$ batch 659). Public LB and Private LB refers to the score in the Kaggle
competition using public test dataset and private test dataset. The score for competition metric is the smaller the better.}
\begin{tabular}{lcccccccc}
\hline
model & $d_{model}$ &  heads & layers & params & CV512 & Public LB & Private LB \\
\hline
T d32 & 192 & 3/6 & 4+12 & 7.57M & 0.9704 & 0.9693 & 0.9698 \\
S d32 & 384 & 6/12 & 4+12 & 29.3M & 0.9671 & 0.9654 & 0.9659 \\
B d32 & 768 & 12/24 & 4+12 & 115.6M  & 0.9642 & 0.9623 & 0.9632 \\
B d64 & 768 & 12/24 & 4+12 & 115.6M  & 0.9645 & 0.9635 & 0.9629 \\
\hline
\end{tabular}
\label{tab:table3}
\end{table}

\begin{table}[H]
\centering
\footnotesize
\caption{Hyper-parameters and score of ISeeCube. ``S-RegA'', ``S-RegB'' and ``S-ClsB'' refers to different ``Small'' models (see section~\ref{sec:headings33}). The score for competition metric is the smaller the better.}
\begin{tabular}{lccccccccc}
\hline
model & $N$ & $d_{model}$ & $d_{ffn}$ & heads & layers & rel pos & max & params & score \\
 & & & & & & buckets & rel pos & & \\
\hline
S-RegA & 196 & 384 & 1536 & 12 & 16 & 32 & 256 & 31.7M & 0.9651 (CV256) \\
S-RegB & 196 & 384 & 1536/1536 & 12/12 & 6+12 & 32/0 & 196/0 & 35.1M & 0.9757 (CV196) \\
S-ClsB & 196 & 384 & 1536/1536 & 12/12 & 6+12 & 32/0 & 196/0 & 43.6M & 1.0095 (CV196) \\
\hline
\end{tabular}
\label{tab:table4}
\end{table}

Model S-RegA follows the learning rate schedule as table~\ref{tab:table5}. Model S-RegB and S-ClsB follows the learning rate schedule as table~\ref{tab:table6}. The reason for epoch 0 in table~\ref{tab:table5} is that we first train our model for 1 epoch to test the performance and we think loading this pre-trained model for further training won't lead to overfitting, considering that (a) the score now is relatively high and (b) there are following 7 epochs with different dataset (see chunk-based method mentioned in section~\ref{sec:headings41}). In order to save time and resources when training, for epoch 0$\sim$48, we use $N=196$ when loading the data (the pulses in an event is 196 after normalization), and for epoch 49$\sim$56, we use $N=256$. In table~\ref{tab:table5} and table~\ref{tab:table6}, ``COMB'' loss means combining two losses by adding them up:
\begin{equation}
\begin{aligned}
COMB &= \Psi + 0.05 \cdot MSE \\
COMB &= \Psi + 0.05 \cdot CE
\end{aligned}
\end{equation}
where $\Psi$ is competition metric in the Kaggle competition. In table~\ref{tab:table5} and table~\ref{tab:table6}, ``div'' and ``div\_final'' are hyper-parameter used in \texttt{Leaner.fit\_one\_cycle} with 1cycle police \cite{smith2019super} in \texttt{fastai}.

\begin{table}[H]
\centering
\footnotesize
\caption{Learning rate schedule of model S-RegA.}
\begin{tabular}{lcccccc}
\hline
epoch & 0 & 1$\sim$8 & 9$\sim$16 & 17$\sim$24 & 25$\sim$32 & 33$\sim$40 \\
\hline
loss & MSE & MSE & MSE & MSE & COMB & COMB \\
lr\_max & 1e-4 & 1e-4 & 1e-4 & 1e-5 & 1e-5 & 0.2e-6 \\
div, div\_final & 25 & 25 & 25 & 25 & 25 & default \\
\hline
\end{tabular}
\label{tab:table5}
\end{table}

\begin{table}[H]
\centering
\footnotesize
\caption{Learning rate schedule of model S-RegB and S-ClsB.}
\begin{tabular}{lcccccccc}
\hline
epoch & 0$\sim$7 & 8$\sim$15 & 16$\sim$23 & 24$\sim$31 & 32$\sim$39 & 40$\sim$47 \\
\hline
loss & MSE/CE & MSE/CE & MSE/CE & COMB & COMB & COMB \\
lr\_max & 1e-5 & 1e-5 & 1e-5 & 1e-5 & 0.5e-5 & 0.35e-5 \\
div, div\_final & 25 & 25 & 25 & 25 & 15 & 10 \\
\hline
\end{tabular}
\label{tab:table6}
\end{table}

\begin{figure}[H]
	\centering
	\subfigcapskip=-5pt
	\subfigure[Previous training epochs with MSE/CE loss.]{
		  \centering
		  \includegraphics[width=0.45\linewidth]{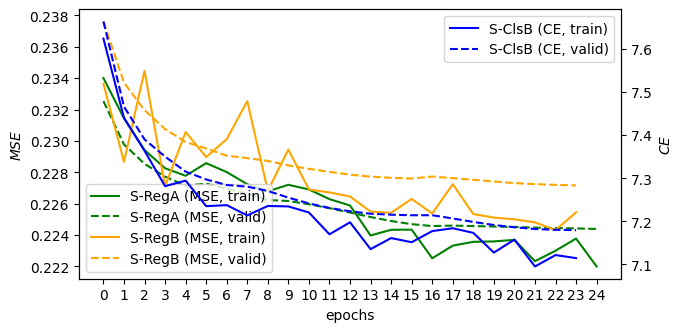}
	}
	\hspace{0.1em}
	\subfigure[Latter training epochs with COMB loss.]{
		  \centering
		  \includegraphics[width=0.45\linewidth]{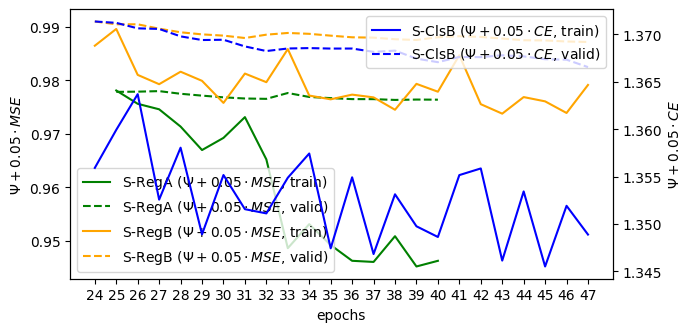}
	}
	\subfigure[Entire validation epochs with competition metric.]{
		  \centering
		  \includegraphics[width=0.9\linewidth]{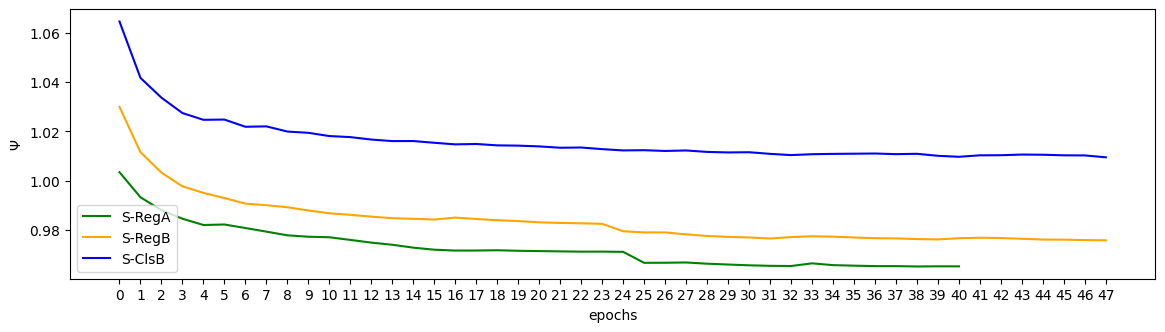}
	}
	\caption{The loss curve of ISeeCube. (a) shows the previous training epochs using MSE/CE loss; (b) shows the latter training epochs using COMB loss; (c) shows the entire validation epochs with competition metric.}
	\label{fig:fig4}
\end{figure}

The loss curve of S-RegA, S-RegB and S-ClsB are shown in figure~\ref{fig:fig4}. As we can see, the training loss curve is flucuating while the validation loss curve is constantly reducing. Also, by replacing MSE loss or CE loss with COMB loss, the loss curve drops in a single epoch.

As shown in table~\ref{tab:table3} and table~\ref{tab:table4}, by adding merely 2.4M parameters, our best model S-RegA outperforms $2^{\textrm{nd}}$ place solution's S d32 model and approaches its B d32 model. Also, compared with CV512, our model uses CV256, which means that our model requires about half the GPU memory when inferencing.

By using voting rule as the ensemble method we reaches a score of 0.9645, which would be $3^{\textrm{rd}}$ place solution in the Kaggle competition. Though as shown in table~\ref{tab:table3}, the score for Public LB and Private LB tend to be better than the score of CV. Also, by traing a B model with roughly 120M parameters and using ensemble, our model could reach state-of-the-art.

\begin{figure}[H]
	\centering
        \includegraphics[width=0.45\linewidth]{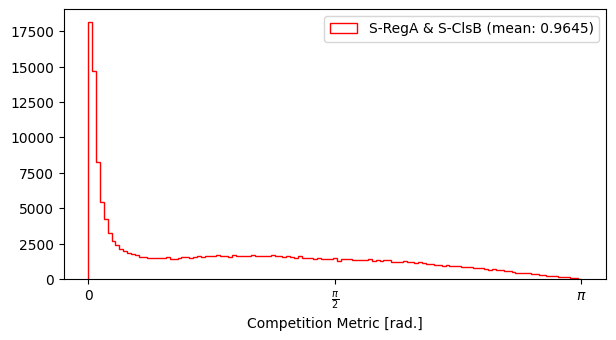}
        %\captionsetup{width=0.85\textwidth}
	\caption{The distribution of competition metric score of batch 655. By using voting rule as the ensemble method, our model reaches a score of 0.9645.}
	\label{fig:fig5}
\end{figure}

\subsection{Azimuthal and Zenithal Distribution}
\label{sec:headings43}

\begin{figure}[H]
	\centering
	\subfigcapskip=-5pt
	\subfigure[Azimuth ($\varphi$) distribution.]{
		  \centering
		  \includegraphics[width=0.9\linewidth]{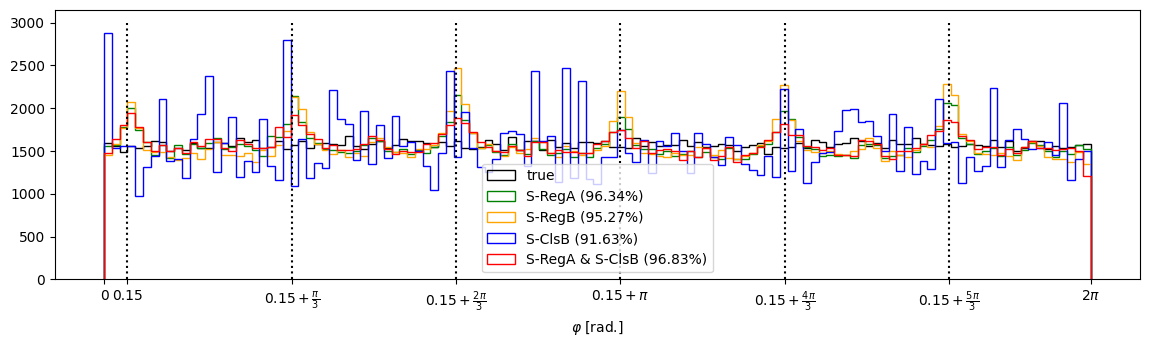}
        }

        \subfigure[Zenith ($\theta$) distribution.]{
		  \centering
		  \includegraphics[width=0.9\linewidth]{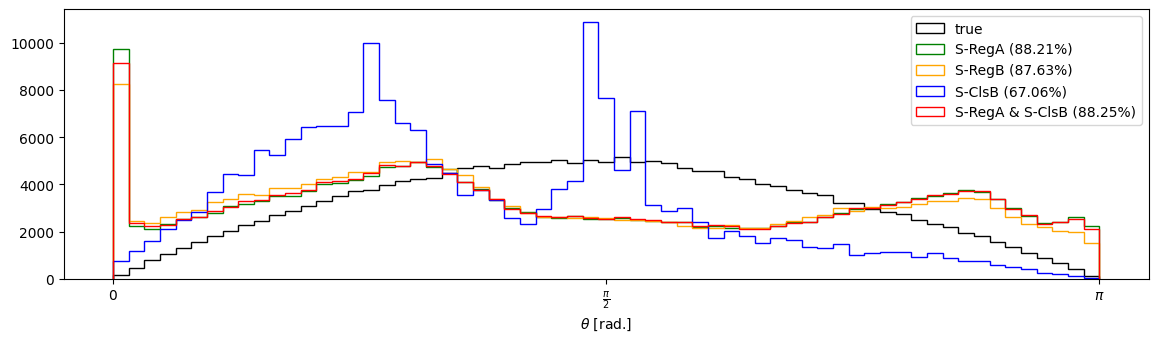}
        }
  
	\subfigure[Azimuthal and zenithal error of ensemble ($\Delta \varphi$ and $\Delta \theta$).]{
		  \centering
		  \includegraphics[width=0.55\linewidth]{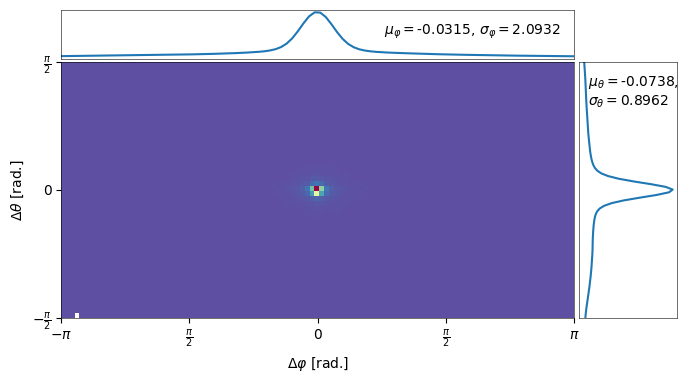}
	}
	\hspace{0.1em}
	\subfigure[Azimuthal and zenithal error of ensemble ($\Delta \varphi$ and $\Delta \theta$), zoomed in.]{
		  \centering
		  \includegraphics[width=0.3\linewidth]{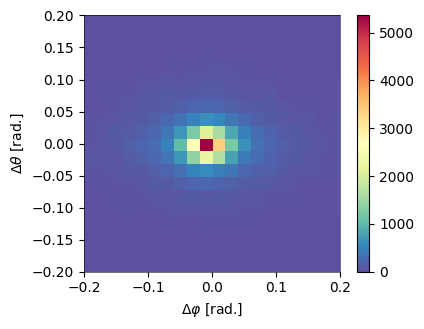}
	}
	\caption{Inference on batch 655. (a) shows the distribution of azimuth; (b) shows the distribution of zenith. The percentage in (a) and (b) refers to overlap ratio between prediction and true value. (c) shows the distribution of azimuthal and zenithal error of the ensemble, while (d) is zoom in of (c).}
	\label{fig:fig6}
\end{figure}

Figure~\ref{fig:fig6} shows the azimuthal and zenithal distribution of the prediction. For azimuthal distribution, we notice an interesting phenomenon that the prediction has 6 peaks corresponding to the shape of IceCube, which means that our model is effected by the nuance of the geometric shape of IceCube. We define overlap ratio:
\begin{equation}
\rho = \frac{\text{overlap area of the prediction and the true value}}{\text{area of the true value}}
\end{equation}
As shown in figure~\ref{fig:fig6} (a), for azimuth prediction, the overlap ratio of both regression model and classification model are over 90\%. It's possible to use overlap ratio as loss function, with bigger batch size when training.

On the contrary, the zenithal distribution is irregular and has lower overlap ratio. The predictions made by regression models tend to be drifted towards $\theta = 0$, while the predictions made by classification models tend to be drifted towards $\theta = \frac{\pi}{3}$ and $\theta = \frac{\pi}{2}$. Regression models have a higher overlap ratio than classification models.

\begin{figure}[H]
	\centering
        \includegraphics[width=0.45\linewidth]{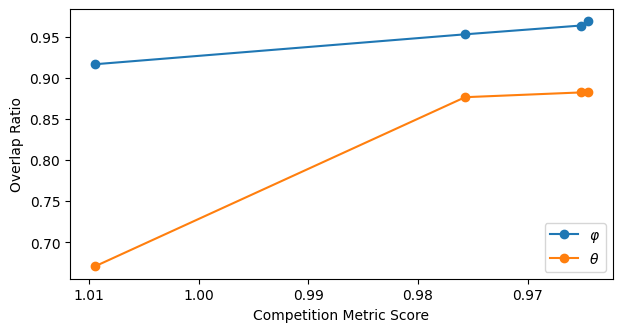}
        %\captionsetup{width=0.85\textwidth}
	\caption{Fitting competition metric score with overlap ratio. The observed data indicates that there is a positive correlation between the overlap ratio and the improvement of competition metric score.} 
	\label{fig:fig7}
\end{figure}

As we can see from table~\ref{tab:table4} and figure~\ref{fig:fig6}, this ratio uniformly reflects the performance of the model. By fitting the score of the competition metric and the overlap ratio with four data points (S-RegA, S-RegB, S-ClsB and ensemble), we can get figure~\ref{fig:fig7}. The curve for azimuth ($\varphi$) is more linear than the curve for zenith ($\theta$), this is most likely due to the fact that azimuth distribution is more uniform than zenith distribution, as shown in figure~\ref{fig:fig6}.

We use some of the most common loss in machine learning to investigate the fundamental models for Neutrino Events Reconstruction. Regression model and Classification model both have their own bias on certain ranges and we are still looking for a unified model. By using ensemble, we can lower the competition metric score, but the distribution of zenith is largely close to regression. Thus ensemble is not the unified model.

\section{Discussion and Conclusion}
\label{sec:headings5}

Neutrino Events Reconstruction has always been crucial for neutrino observatories such as IceCube and KM3NeT/ORCA, which traditionally uses CNNs or GNNs. In the Kaggle competition, many solutions use Transformers. And it is shown that a pure Transformer model performs really well for Neutrino Events Reconstruction.

Compared with GNN integrated with Transformer or an ensemble of several different models including Transformer, GNN and RNN, a pure Transformer structure is easier for refining, maintaining and testing. And integrating other models with this model would be much easier. In this spirit, we build ISeeCube based on the $2^{\textrm{nd}}$ place solution, by replacing BEiT v2 with $\texttt{TorchScale}$ (the backbone of BEiT-3). We also uses Register Tokens, improving the performance while only slightly increase the computation cost. The lines of code drop sharply and a lot of new methods could be tested by simply adjusting the config or using other models as the first layer. When having relatively same amount of total parameters, our model outperforms the $2^{\textrm{nd}}$ place solution.

We use rather primitive loss functions such as MSE loss and CE loss in order to explore the fundamental models for Neutrino Events Reconstruction. We then investigated how azimuthal and zenithal distribution would be effected by the choice of regression model or classification model. We show that Transformer, especially Vision Transformer has the potential to learn the information of graphs and time series in neutrino events. Future upgrades on Transformer Encoder or Vision Transformer can be used for Neutrino Events Reconstruction.

% Bibliography

%% [A] Recommended: using JHEP.bst file
%% \bibliographystyle{JHEP}
%% \bibliography{biblio.bib}

%% or
%% [B] Manual formatting (see below)
%% (i) We suggest to always provide author, title and journal data or doi:
%% in short all the informations that clearly identify a document.
%% (ii) please avoid comments such as "For a review'', "For some examples",
%% "and references therein" or move them in the text. In general, please leave only references in the bibliography and move all
%% accessory text in footnotes.
%% (iii) Also, please have only one work for each \bibitem.

\end{document}